\documentclass[10pt,letterpaper]{article} 
\usepackage[english]{babel}
\newcommand{\s}{\\[1ex]}
\newcommand{\goesto}{\scriptstyle\rightarrow}
\newcommand{\beq}{\begin{equation}} 
\newcommand{\eeq}{\end{equation}}
\newcommand{\lbl}{\label}
\newcommand{\q}{\quad}

\newcommand{\re}[1]{(\ref{#1})} 
\begin{document} 
\title{Comment on {\em Phys. Rev. D} {\bf 60} 084017\\
``Classical self-force'' by F. Rohrlich\\
}
\author{Stephen Parrott \\ 
2575 Bowers Rd.\\
Gardnerville, NV 89410\\
Email:  spardr@netscape.net
}
\maketitle
\noindent 
\s
\noindent
\begin{abstract}
F.\ Rohrlich has recently published two papers, including 
the paper under review, advocating 
a particular delay-differential equation as 
an approximate equation of motion for classical charged particles,  
which he characterizes as providing a 
``fully acceptable classical electrodynamics''. 
This Comment notes some mathematical and physical problems
with this equation.
It points out that most of the claims of these papers are unproved,  
while some appear to be false as stated. 
\end{abstract}

\section{Introduction}
The correct equation to describe the motion of a charged particle
in flat spacetime (Minkowski space)
has long been a matter of speculation and controversy.
The most-mentioned candidate has been the Lorentz-Dirac equation,
written here for units in which light has unit velocity
and metric tensor of signature $[+,-,-,-]$ :
\beq
\lbl{ldeq}
m \frac{du^i}{d\tau} = q {F^i}_\alpha u^\alpha + \frac{2}{3} q^2
\left[\frac{d^2 u^i}{d\tau^2} + \frac{du^\alpha}{d\tau}
\frac{du_\alpha}{d\tau} u^i 
\right]
\eeq
This is written in traditional tensor notation with repeated indices 
summed and emphasized in Greek;
$u = u^i$ denotes the particle's four-velocity, $m$ and $q$ its mass
and charge, respectively, $\tau$ its proper time,
and $F_{ij}$ the (antisymmetric) tensor describing an external electromagnetic
field driving the motion.  

However, many objections have been raised to this equation.
Among them are the existence of ``runaway'' solutions
for which the acceleration increases exponentially with proper time
even when the external field asymptotically vanishes.
Indeed, in some physically reasonable situations, 
{\em all} solutions are runaway 
\cite{eliezer}.
Even in favorable cases in which non-runaway solutions exist,
they may exhibit ``preacceleration'', in which the particle
begins to accelerate {\em before} the external field is applied.

F.\ Rohrlich \cite{rohrlichpr} \cite{rohrlichamj} 
has recently advocated a new (approximate) equation of motion which
\cite{rohrlichpr} says ``is to replace'' the Lorentz-Dirac equation. 
This new equation is a delay-differential equation, given below as 
\re{rohreq1},
which we shall call the ``DD equation''.

Rohrlich claims without proof that the ``great virtue'' of the DD equation 
is that it 
has ``no pathological solutions'' \cite{rohrlichpr},
in particular ``no  runaways and no preaccelerations'' \cite{rohrlichamj}. 
We shall explain why these claims are at best optimistic and 
at worst false.
Detailed proofs for all the statements we shall make
can be found in the ArXiv paper www.arxiv.org/gr-qc/0205065,
which was judged mathematically too technical for this journal.
\section{Initial conditions for the DD equation}
The DD equation, in the above notation, is:
\beq
\lbl{rohreq1}
m_1 \frac{du^i}{d\tau} = 
f^i(\tau) +  
m_2 [ u^i(\tau-\tau_1) - u^\alpha (\tau - \tau_1) u_\alpha (\tau) u^i(\tau) ] 
\q.
\eeq 
\newcounter{rohreqonenum} 
\setcounter{rohreqonenum}{\value{equation}} 
Here $f(\tau) = f^i(\tau) $ is a four-force orthogonal to $ u(\tau)$, 
$m_1$ and $m_2$ are positive  
parameters associated with his motivation
of the right side as an approximation to the self-force on a 
spherical surface charge, and the delay $\tau_1$  is a positive parameter.
(The sign of the second term in brackets differs from Rohrlich's because
our metric is opposite in sign to his.)  
Below we shall omit the indices on four-vectors like $u$ 
where no confusion is likely.

This is easier to discuss if we write it in simpler but more abstract
form as  
\beq
\lbl{delayeq}
\frac{du}{d\tau} = g(u(\tau), u(\tau - \tau_1), \tau)
\q,
\eeq
where $g$ is a given function.  
The $g$ corresponding to  \re{rohreq1} is 
\beq
\lbl{DDg}
g(y,w, \tau) := \frac{f(\tau)}{m_1} + 
\frac{m_2}{m_1}
[w -     ( w^\alpha y_\alpha)y ]
\q,
\eeq
but the precise form of $g$ 
will not play an important role in the discussion. 

The possible dependence of $g(y,w,\tau)$ on $w$ in \re{delayeq}
is what makes \re{delayeq} a {\em delay}-differential equation
rather than an ordinary differential equation.
If $g(y,w, \tau)$ happens to be independent of its second argument $w$, then
\re{delayeq} becomes an ordinary differential equation whose solutions
are uniquely determined by prescribing the values of $u^i (\tau)$
at some initial time $\tau_0$, say $u^i(\tau_0) = c^i$ where
$c^i$ is given.  Thus in this case, the solution space
is a manifold of dimension 4, parametrized by choice of initial conditions
$c^i$.

In the case of a hypothetical equation like 
$du/d\tau = g(u(\tau), \tau)$ for the four-velocity $u$ of a particle,
one has to adjoin the condition $u^\alpha u_\alpha = 1$, which must
be satisfied by any four-velocity.  This reduces the dimension of
the solution manifold from 4 to 3; we omit the details. 
And this is consistent with physical expectation---there
ought to be precisely one solution for each initial three-velocity.
(When $u^i(\tau) := dz^i/d\tau$ is integrated to obtain the particle's wordline
$z^i(\tau) \in R^4$,
four more arbitary constants appear.)

But if $g(y,w,\tau)$ is {\em not} independent of $w$,
as in the DD equation \re{rohreq1},
the nature of the solution manifold changes drastically.
Barring unlikely degeneracies, 
it is expected to be 
{\em infinite}-dimensional, and this is indeed the case for
the DD equation.
This is because to obtain a unique solution to \re{delayeq},
we need to specify not just an initial value $u(\tau_0)$ 
but the values of $u(\tau)$ over an {\em entire interval} 
of length $\tau_1$, say
$[\tau_0- \tau_1, \tau_0 ]$, subject to the  consistency condition
\beq
\lbl{consistency}
\frac{du}{d\tau}(\tau_0 ) = g(u(\tau_0 ), 
u(\tau_0 - \tau_1), \tau_0)
\q.
\eeq
For brevity, when we speak of a ``specification of $u(\tau)$
on $[\tau_0 - \tau_1, \tau_0 ]$'', we shall understand this to
include the consistency condition. It is also understood 
that the specification of $u$ is to make $u$ a possible four-velocity, 
i.e., $u^\alpha(\tau) u_\alpha(\tau) = 1$ and $u^0 \geq 1$
for $\tau_0 - \tau_1 \leq \tau \leq \tau_0$).

To understand this intuitively, 
imagine how one could solve \re{delayeq} given the values $u(\tau)$
on an interval $[\tau_0 - \tau_1, \tau_0 ]$.   
First substitute these values in  \re{delayeq} for 
$\tau_0  \leq \tau \leq \tau_0 + \tau_1$, thus obtaining an {\em ordinary}
differential equation for $u(\tau)$ on 
$[\tau_0 , \tau_0 + \tau_1]$.
Solving this equation gives the solution $u(\tau)$, which was initially
known only on $[\tau_0 - \tau_1, \tau_0 ]$, on the new
interval $[\tau_0 , \tau_0 + \tau_1]$.  Repeating the process
gives a unique solution on $[\tau_0, \infty)$.  

Solving \re{delayeq} backwards in time, i.e., for $\tau < \tau_0  $, 
requires additional hypotheses on $g$.  
Appendix 5 of \cite{parrottpaper} shows how to solve the DD equation
backward in time.  The same method 
works for a non-relativistic approximation \re{nonrelDD} to the DD equation
to be discussed later.  
Henceforth, we specialize the discussion
to these two equations.

In summary, to each specification of $u(\tau)$ on a given interval
$[\tau_0 - \tau_1, \tau_0 ]$ of length $\tau_1$, 
there corresponds a unique solution of 
the DD equation \re{rohreq1} 
on $(- \infty, \infty)$, and the same is true for its nonrelativistic
approximation \re{nonrelDD}.
We shall call such a specification a ``generalized initial condition''.

This implies that the DD equation 
has far more solutions than physically expected---we 
cannot find a unique solution
satisfying an ordinary initial condition 
${\bf u} (\tau_0) = {\bf u}_0$, where ${\bf u} := (u^1, u^2, u^3)$ 
denotes the space part of $ u = (u^0, u^1, u^2, u^3)$.   
In any practical situation, we will not know which of the infinitely
many solutions satisfying 
${\bf u} (\tau_0) = {\bf u}_0$
will accurately describe the particle's motion.

One might counter that in principle, 
we could observe the particle's behavior over a time interval of
length $\tau_1$ and use that as a generalized initial condition.
But in Rohrlich's motivation of the DD equation,
the delay $\tau_1$ is the particle's diameter, the time it takes
for light to cross the particle. Such an observation seems
hopeless in practice, and arguably impossible in principle.

To obtain a physically reasonable or useful theory, additional 
conditions forcing the solution manifold down to 3 dimensions 
(or at the very least, to a finite number of dimensions)
seem necessary.%
\footnote{The solution manifold is physically expected
to be three-dimensional in a context in which the external force 
$f(\tau)$ is regarded as given, {\em a priori}, as a function
of proper time $\tau$.  In other contexts, such as a force 
caused by a space-dependent electric field, one has to
rewrite the equation of motion  as an equation for the worldline
instead of the four-velocity, in which case the solution manifold 
would be expected to be seven-dimensional.
\s
We remark in passing that
the solution manifold of the Lorentz-Dirac equation is also
regarded as too large because it is second order in $u$,
so that to obtain a solution, one needs to specify an initial
acceleration along with the usual initial conditions of
initial position and velocity.
Some of the pathologies of that equation arise from 
this circumstance.
} 
Such conditions should  be provided before proponents
of the DD equation can reasonably propose it as
a sensible replacement 
for the Lorentz-Dirac equation.

There is one special case in which there is a unique
reasonable
choice of such auxiliary conditions---the case in which the force $f$
in \re{rohreq1} is applied for only a finite time.  
This will be discussed in the next section.
But no such conditions are obvious or have been proposed for 
general forces.  A discussion of some of the mathematical
considerations is given in Appendix 1 of \cite{parrottpaper}. 

\section{The case of a force applied for only a finite time}
Suppose that the force $f^i(\tau)$ in the DD equation is nonzero for only 
a finite time, say  $\tau_0 < \tau < \tau_2$. 
Then a solution $u$ of the DD equation \re{rohreq1} is called
{\em preaccelerative} if $du/d\tau$ does not vanish identically
on $(-\infty, \tau_0]$, and {\em postaccelerative} if $du/d\tau$
does not vanish identically on $[\tau_2, \infty)$.
A more general definition of preacceleration 
applicable to forces active for all time (as are most forces in nature)
is given in \cite{parrottpaper}; the simpler definition just given
was introduced for expositional simplicity. 

Preaccelerative solutions are generally considered physically
unreasonable, and postaccelerative solutions also
seem physically questionable.  
For example,
suppose we are sitting in a room shielded from electromagnetic fields
watching a beam of identical charged particles shoot in the window.
It might seem strange if some of the particles 
speeded up, while others slowed down, for no apparent reason,
according to their past histories.  
We shall see that this is what the DD equation predicts.

The solution $u$ will be called ``runaway in the past'',
or ``past runaway'' if the acceleration $du/d\tau$ does not approach zero as 
$\tau \goesto - \infty$, and ``future runaway'' if 
$du/d\tau$ does not approach zero as $\tau \goesto \infty$. 

These are not quite the usual definitions of ``runaway''---more typically,
``runaway'' is used in the sense of exponential increase.
We use the weaker definition because it is only under this definition
that assertions given below concerning runaway solutions have been proved;
nevertheless, it seems probable that these assertions would also
hold under the more usual definitions. 

In the special situation under consideration (a force applied for
only a finite time), past runaways are automatically preaccelerative,
so a generalized initial
condition which eliminates preacceleration also guarantees that 
the solution cannot be runaway in the past. 
However, for the more general definitions of \cite{parrottpaper},
this may not be true.

Rohrlich \cite{rohrlichamj} claims without proof that the DD equation
has
\begin{quote} 
``no unphysical solution, no runaways, and no preaccelerations.''
\end{quote}
We shall indicate why these claims are at best optimistic,
and at worst false. 

The claim of ``no preaccelerations'' is false as stated.
Suppose $f^i (\tau) = 0$ for $\tau \leq 0$. 
Then the generalized initial condition necessary to prevent preacceleration
is 
\beq
\lbl{initcond}
u(\tau) \equiv \mbox{constant} \q \mbox{for $ -\tau_1 \leq \tau \leq 0$.} 
\eeq 
This condition is obviously necessary (because if there is no preacceleration,
$du/d\tau$ vanishes identically on $(-\infty, 0]$ by definition), 
and we'll see below that  it is also sufficient.

Given a solution $u$ of the DD equation on $[-\tau_1, 0]$  
it is a routine exercise to solve
for $u$ on $[-2\tau_1, -\tau_1]$, and, by iteration, 
on any interval $[-(n+1) \tau_1 , -n\tau_1 ]$ 
for any given positive integer $n$.
The analysis is particularly easy 
for the case of motion in one space dimension,
and is explicitly done in \cite{parrottpaper}.
For simplicity, we restrict the rest of the discussion
to this case. 

The result when $u$ satisfies \re{initcond} is that
$u$ is constant on $(-\infty, 0]$.
When $u$ does not satisfy \re{initcond}, 
a closed form solution on $(-\infty, 0]$ may be difficult to obtain, 
but nevertheless one can show \cite{parrottpaper} that
$u$ is both preaccelerative (obviously) and runaway in the past
(not obvious, but provable).  

In summary, \re{initcond} is the {\em unique} generalized initial condition
to prevent preacceleration and past runaways.
This shows that the claim \cite{rohrlichamj} that ``[the DD equation has]
 no preaccelerations''
is false as stated.

In this special situation, this is not so serious---we can
repair the claim of no preaccelerations 
by simply imposing the  correct generalized initial condition \re{initcond}. 
Unfortunately, this repair is not available for forces which
do not vanish for large negative times, 
and there seems no obvious alternative repair. 

More serious, perhaps, is that 
it is not generally possible to eliminate preacceleration and postacceleration
simultaneously.
The proof, given in \cite{parrottpaper},
is not difficult.
The idea is that preacceleration is eliminated by solving forward in time
from a condition \re{initcond} of zero acceleration in the past,
while postaccelertion is eliminated by solving backward in time from
a condition of zero acceleration in the future, and there is nothing 
in the mathematics forcing these two solution methods to produce 
the same solution. 

For example, for the DD equation (for motion in one space dimension)
with a force $f(\tau)$ which vanishes off the interval 
$0 \leq \tau \leq \tau_1$, 
one can show that the solution is {\em always} either preaccelerative
or postaccelerative except in the special case of identically 
vanishing force.  
This example may be physically uninteresting (because the delay
parameter is expected to be so small),
but it well illustrates 
how elimination of preacceleration and postacceleration
are basically independent and usually incompatible. 

For forces supported on larger intervals, solutions which are
not preaccelerative may happen to exhibit no postacceleration,  
but this requires the force $f$ to be exquisitely ``fine-tuned''.  
For most forces, solutions which are preaccelerative will exhibit
postacceleration.  
Thus if one considers postaccelerative solutions ``unphysical'',
it is hard to see how
the claim of ``no unphysical solution'' could be repaired. 

The claim of no runaways turns out to be true for the special case
in which
\begin{description}
\item[{\rm (a)}]
The particle's motion and the applied 3-force are in one spatial
dimension (e.g., the $x$-axis); 
\item[{\rm(b)}]
the force is applied for only a finite time; and 
\item[{\rm(c)}]
the correct initial condition \re{initcond} is imposed.  
\end{description}
We have already noted the nonexistence of past runaways under these
hypotheses.
However, it turned out to be surprisingly difficult 
to prove the nonexistence of future runaways. 
(This is the only result of real mathematical substance 
in \cite{parrottpaper}.) 

It does not look routine to extend the proof to general three-dimensional motion. 
Thus the claim of ``no runaways'' for general three-dimensional motion
remains to be proved, even for the special case of a force applied for
only a finite time. 
And, all the above claims of \cite{rohrlichpr} and \cite{rohrlichamj}
not covered by the above remarks
remain to be proved for general forces. 
\section{The analysis of a nonrelativistic version of the DD equation
by Moniz and Sharp}

The only evidence offered by \cite{rohrlichpr} or \cite{rohrlichamj}
for its claims of no preaccelerations, runaways, or unphysical solutions
is  a 1977 paper of Moniz and Sharp \cite{ms} analyzing the following non-relativistic
version of the DD equation:
\newcommand{\bu}{{\bf v}}
\newcommand{\bbf}{{\bf h}}
\newcommand{\ba}{{\bf u}_0} 
\beq
\lbl{nonrelDD}
\frac{d\bu}{dt} = \bbf(t) - b [ \bu (t - \tau_1) - \bu(t)]
\q,
\eeq
Here $\bu (t)$ represents the particle's three-dimensional velocity
at time $t$, $\bbf$ is a force-like term  (a three-dimensional force
divided by certain constants),
and $b$ is a constant. 
This can be obtained from the DD equation by 
replacing the right side by an approximation which is at most 
of order $|\bu|$ (i.e., expand in power series and delete terms
of quadratic or higher order in the velocity), rearranging algebraically,
and renaming the term involving the force.

Note that this is what one might call a ``linear'' delay-differential
equation, and so is vastly simpler than the DD equation \re{rohreq1}. 
The linearity of \re{nonrelDD} 
enables Moniz and Sharp to analyze it using Fourier transforms,
a technique which is difficult to adapt to nonlinear equations.

Moniz and Sharp \cite{ms} make claims similar to the above-cited
claims of \cite{rohrlichamj} 
regarding nonexistence of preaccelerations and runaways:
\begin{quote}
``Summarizing, we have found that including the effects of radiation
reaction on a charged spherical shell results neither in runaway
behavior nor in preacceleration if the charge radius 
of the shell $L > c\tau \ldots$'' 
\end{quote}
(Their condition $L > c\tau$ translates in our notation to a condition
that the delay parameter $\tau_1$ be at least as large as a certain
positive constant, whose value is not relevant here.) 
But the truth of these claims does not imply the truth 
of similar claims for the more complicated, nonlinear, DD equation. 

Moreover, even Moniz and Sharp's claims may not be entirely 
true and, and those that may be true are not entirely proved
in their paper \cite{ms}.  
Appendix 3 of \cite{parrottpaper} analyzes precisely what their
mathematics does prove. 

It points out a serious flaw in 
their proof of the nonexistence of runaways,
and it looks unlikely that this proof can be repaired.  
It also notes that while
they do write down a formal%
\footnote{``Formal'' is used in the mathematical sense of  
``algebraic'', with the implication that analytical 
subtleties are not addressed.
For example,
they write down an expression (containing singularities) 
for the Fourier transform of the solution, but do not  consider
the question of whether there actually is a solution
with this Fourier transform.} 
 Fourier transform of a solution
which eliminates preaccelerations, their proof is incomplete
because the Fourier transform which they furnish for their formal solution  
is {\em unique}, leaving no room to satisfy arbitrary initial
conditions ${\bu}(t_0) = \bu_0$.  
It seems possible that this proof could be completed, but it seems unlikely 
that its techniques could be extended to apply to the 
nonlinear DD equation \re{rohreq1}. 
\section{Appendix 1:  Referees' reports} 
\begin{description}
\item[Note:]
In the original version of this paper, this section contained 
all the referees' reports verbatim.  
That version was rejected by the arXiv because of concerns
that including the reports verbatim (rather than paraphrased,
which is undisputedly acceptable)
might somehow infringe some copyright.  
(This is entirely hypothetical because no one has asserted any 
rights of copyright to the referees' reports, nor made any objection
to their inclusion in the arXiv.) 
\s
The reports were quoted verbatim only to avoid any possible distortion
of the referees' meanings.  Given this and 
given that the referees' reports have absolutely no commercial value,
I believe that the verbatim quotes are permitted under the ``Fair Use" 
section of the copyright law (Section 107), which makes an exception
for scholarly use:
\begin{quote}
``$\ldots$ the fair use of a copyrighted
work
for purposes such as $\ldots$ scholarship, or research, 
is not an infringement of copyright. In determining
whether the use made of a work in any particular case is a fair use 
the factors
to be considered shall include
\begin{enumerate}
\item
the purpose and character of the use, including whether such use is of a
commercial nature or is for nonprofit educational purposes;
\item 
the nature of the copyrighted work;
\item
the amount and substantiality of the portion used in relation to the copyrighted
work as a whole; and
\item
the effect of the use upon the potential market for or 
value of the copyrighted work.'' 
\end{enumerate}
\end{quote} 
Nevertheless, to avoid an unpleasant dispute,
I rewrote this section so that the verbatim reports are replaced
by paraphrased ones. 
So long as I am convinced that the ``Fair Use'' section applies
to this case, 
I will furnish by email or other means the original version
containing the verbatim reports to anyone who requests one.
Visit www.math.umb.edu/$\sim$sp for further information.
\end{description} 
The above ``Comment'' paper was submitted to  
Physical\ Review\ D (PRD). 
It was rejected.
The chronology of the rejection is as follows.

\begin{enumerate}
\item
PRD invited Professor Rohrlich to submit his 
views on my ``Comment'' paper.  These are paraphrased below. 
Paraphrases are allowed to contain brief quotes.  The italization in
the quotes is his. 
\begin{description}
\item[Paraphrase of Professor Rohrlich's report:]
Rohrlich's report begins with a brief summary of his version of the history 
of the DD equation.  He attributes it to Caldirola and to Yaghjian 
and specifically states that it was {\em not} first proposed by Rohrlich.

He then expresses disappointment
that the DD equation does have preaccelerative solutions,
contrary to his previous expectation.  He goes on to say 
that nevertheless, this is ``physically irrelevant''
because 
\begin{quote}
``{\em the preaccelereation problem has been solved} 
 for the point particle case''.  Therefore there is no longer
any need to repair it by using [the DD equation].''
\end{quote}
Then he states that ``a physically meaningful differential equation
of motion for a charged
particle {\em now does exist} for classical physics''.
Later in the review he identifies this equation as the Landau-Lifschitz
equation.  

He goes on to claim that this 
\begin{quote}
``$\ldots$ was {\em first justified mathematically}
five years ago by H. Spohn, Europhysics Letters 50, 287-292 (2000).
He showed that the Lorentz-Dirac equation has in its solution 
space a critical manifold to which all physical solutions must be 
restricted.  Within the validity domain of classical physics,
this restriction can be accomplished by approximating the Lorentz-Dirac
equation by the Landau-Lifshitz equation $\ldots$.
\end{quote}
He concludes that the present Comment paper, ``though correct'',
is ``no longer of physical interest'' and therefore should not be published.
\end{description}
\item
I submitted a reply which is reprinted in Appendix 2.
\item
PRD then sent the ``Comment'' paper to an anonymous referee.
I assume that Professor Rohrlich's comments were also forwarded to
this referee because his report refers in passing to Spohn's work
without giving a specific reference.

The referee's short report states that no existing experiment can distinguish
between various proposals for an equation of  motion for classical charged
particles. Given this, he says that ``it does not make much sense
to discuss mathematical subtleties of the different equations 
for accelerated charges''. 
The report concludes  that the Comment paper is 
therefore ``not of physical interest''
and should not be published in Phys. Rev. D.

Both the anonymous referee and the Editor's rejection letter
entirely ignored Rohrlich's 
claims to the effect 
that Spohn's work solves the problems which the DD equation was supposed to  
solve, thereby making the DD equation 
``no longer of physical interest'',  
Because of this,
I think it fair to say that the paper was rejected
solely because the editors accept the anonymous referee's opinion
that the entire subject of equations of motion for classical charged
particles is ``not of physical interest''. 
\end{enumerate}
\section{Appendix 2:  Author's reply to Professor Rohrlich's report} 
The following was sent to PRD in response to Professor Rohrlich's report.
Presumably, PRD sent it to the anonymous referee. 
\begin{center}
\large
\bf
Author's reply\\
\normalsize
\bf
to\\
\large
\bf
the report of F.\ Rohrlich\\ 
\end{center}
\subsection{Introduction}
The Lorentz-Dirac equation is an equation of motion for charged particles
which was originally derived in order to ensure conservation
of energy-momentum of
a charged particle together with that of
the electromagnetic field.
Its derivation requires the controversial principle of mass renormalization,
but otherwise is mathematically rigorous and free of approximations.
Because of the rigor of its derivation, for about sixty years
it has been the most-mentioned candidate for such an equation of motion. 

The reason it is only a candidate, rather than an accepted physical
principle, is that it has consequences so strange 
that no one will admit to believing them.
Some of these are:
\begin{enumerate}
\item
It admits solutions which are ``runaway'' in the sense that
the particle's energy increases exponentially with proper time
even when the external field driving the motion is only 
applied for a finite time.  In favorable cases, runaway solutions
may be eliminated by appropriate choice of initial conditions,
but in some physically reasonable situations, {\em all}
solutions are runaway.  
\s
For instance, this is the case for an electron which starts at rest
and moves radially in the Coulomb field of a stationary proton, 
a fact proved by Eliezer in 1943 \cite{eliezer}.
He showed that, according to the Lorentz-Dirac equation,
the electron is not attracted to a collision with the proton
as expected.  Moreover, in the asymptotic future, the electron flees outward from
the proton with exponentially increasing acceleration.
In particular, {\em this contradicts the basic notion that unlike
charges attract} each other, a conclusion which no one seems to believe. 
\item
If a charged particle moves in an electromagnetic field applied
for only a finite time, some solutions of the Lorentz-Dirac equation
are ``preaccelerative'', meaning that the particle begins to accelerate
{\em before} the field is turned on.  This violates basic ideas
of causality.
\s
Sometimes, it is possible to choose initial conditions such that
the solution is not preaccelerative.  However, it is not generally
possible to simultaneously eliminate both preacceleration and runaways.
In ``generic'' situations, eliminating one guarantees the other. 
\item
There are also ``postaccelerative'' solutions for which the particle
accelerates {\em after} the field is turned off.
Indeed, runaway solutions are automatically postaccelerative 
for a force applied for only a finite time.  
\end{enumerate}

These do not exhaust the ``unphysical'' consequences of the Lorentz-Dirac
equation, but they are all we need to consider here.  
Because the Lorentz-Dirac equation has so many unbelievable consequences, 
many authors have proposed replacements for it.   

F.\ Rohrlich's {\em Phys. Rev. D} (PRD) paper 
``Classical self-force'' \cite{rohrlichpr}
proposes a particular equation of motion as a replacement for the Lorentz-Dirac
equation.  My paper calls this equation 
the ``DD equation''.%
\footnote{An earlier version called the DD equation 
``Rohrlich's equation'' because he was the first to state it in the 
literature in full generality, 
but I renamed it after he objected.
}
Rohrlich's PRD paper ``Classical self-force'' 
claims that the DD equation has ``no pathological
solutions''.  A similar paper of Rohrlich in another journal \cite{rohrlichamj}
makes clear that this means in particular 
``no runaways and no preaccelerations''.
These claims were presented without proof as facts,
and without any language cautioning the reader 
that they were merely conjectures. 

My ``Comment'' paper points out that some of these claims are false
as stated, but can be made true in special situations by adding
additional hypotheses.  It points out that 
it might be possible to reformulate some of these conjectures
so that they become true, but that much nontrivial mathematical
work would be necessary to put them a sound mathematical
footing.  

	It notes that  both preacceleration and postacceleration
cannot, in general, be simultaneously eliminated---setting the initial
conditions to eliminate one of these usually prevents eliminating the other.
The situation is similar to the impossibility of eliminating, in general,
both preaccelerative and runaway solutions to the Lorentz-Dirac equation.
\subsection{Professor Rohrlich's latest objection}
A more lengthy paper of which this ``Comment'' paper is a summary,
an earlier version of \cite{parrottpaper}, 
was submitted to PRD in November of 2003.  
Professor Rohrlich served as an identified referee and delivered
a report in December, 2003.
His report made the following objections to the paper, 
and only these:
\begin{enumerate}
\item
The paper called the equation ``Rohrlich's equation'', for reasons
discussed in the introduction to that paper.  He objected to this name,
instead attributing the equation to Caldirola and to Yaghjian.
I had sent him a copy of the paper over a year previously,
but had received no comment, and in particular no objection to
the equation's name.  
\s
In deference to his wishes, I 
renamed the equation the ``DD equation''
in subsequent versions of the paper.  
I thought that this had settled the matter,
but for reasons unclear to me, 
he brings it up again in his latest objection to the ``Comment'' paper,  
so perhaps I should say something about it.
\s
Yaghjian did propose basically the same equation in nonrelativistic
notation, but he presented it only as an uncontrolled approximation
to the Lorentz-Dirac equation without comment on the domain of validity
of the approximation.  By contrast, Rohrlich advocated it 
as a providing a ``fully acceptable classical nonrelativistic dynamics''
\cite{rohrlichamj} which was ``to replace'' \cite{rohrlichpr}
the Lorentz-Dirac equation.  
Because of this, and because Rohrlich was the first to state it in print
in full generality, I originally called it ``Rohrlich's equation''. 
\s
The attribution to Caldirola is simply wrong.
The equation proposed by Caldirola was of a mathematically different kind,
a difference equation involving no derivatives, whereas the DD equation is 
a delay-differential equation.
Solutions of Caldirola's equation are not necessarily solutions of the
DD equation.  There is no reason to think there is any relation
between solutions of Caldirola's equation and solutions of the DD equation.
\item
Professor Rohrlich's report objected, incorrectly, that 
the paper's analysis was physically irrelevant
because the paper replaces the DD equation as presented by Rohrlich
by a superficially different, but essentially equivalent equation, 
in which certain redundant parameters in Rohrlich's version
are collected together as one.  
\end{enumerate}
Objection 2 was the {\em only} substantive objection in that report.

His latest report on the ``Comment'' paper indicates that he has
abandoned Objection 2, does not question the conclusions of the
paper, and has given up   
on the DD equation as 
a likely replacement for the Lorentz-Dirac equation.  
He has made this the basis of the following new objection,  
in which the notes in brackets are mine and the italics his:
\begin{quote}
``Since the publication of the criticized papers 
[presumably he means his papers \cite{rohrlichamj} and \cite{rohrlichpr}],
{\em the preacceleration problem has been solved} for the point
particle case.  Therefor, there is no longer any need trying 
to repair it by using a finite radius.''
\begin{quote}
[Note:  The Lorentz-Dirac equation and DD equations are both mathematically
equations of motion for a moving point.  The moving point could
be visualized a a point particle, or as the ``center'' of an extended 
particle.  Rohrlich \cite{rohrlichpr} \cite{rohrlichamj} 
motivated the DD equation by considering the
particle as a charged sphere with a nonzero radius.] 
\end{quote} 
``A physically meaningful differential equation of motion $\ldots$
for a charged particle {\em now does exist} for classical physics.
It has no pre- (or post-) acceleration solutions.''
\s
That result was {\em first justified mathematically} five years ago
by H. Spohn [\cite{spohn}]$\ldots$.  He showed that the Lorentz-Dirac equation
has in its solution space a critical manifold
to which all physical solutions must be restricted. 
Within the validity domain of classical physics, this restriction
can be accomplished by approximating the Lorentz-Dirac equation
by the Landau-Lifshitz equation of 1962.  $\ldots$''
\s 
``The paper under review, though correct, is therefore no longer
of physical interest.  My recommendation is therefore not to
publish this paper.''  
\end{quote}

I don't agree that the cited work of Spohn \cite{spohn}
solves {\em any} of the problems
with the Lorentz-Dirac equation mentioned in the introduction.
Rohrlich's summary of its results is inaccurate 
and highly misleading.%
\footnote{
Spohn's paper \cite{spohn}
 does {\em not} ``show'' that physical solutions lie on the so-called
``critical manifold---this is merely an unproved claim in a sketchily 
written work  
which contains significant errors.
Even if this claim turns out to be true, it would imply nothing about
situations in which there are {\em no} ``physical'' solutions
(e.g., when all solutions are runaway, as in the conclusion to Eliezer's
theorem). 
\s
Spohn \cite{spohn} does {\em not} show that ``within the domain of classical
physics'', the Landau-Lifshitz equation adequately approximates
the Lorentz-Dirac equation on the critical manifold.
Spohn's approximation is uncontrolled---he gives no rigorous
estimates of the degree of approximation.  
}
 
Spohn's paper \cite{spohn} does not even mention preacceleration.
It might give a casual reader the impression that it somehow
solves the problem of runaway solutions, but a careful mathematical
reading does not support any such conclusion.
And it certainly doesn't negate Eliezer's Theorem stating physically reasonable
circumstances in which the Lorentz-Dirac equation implies
that unlike charges repel each other.

What it does do is give a mathematically 
questionable motivation for an approximate equation
of motion proposed over forty years ago by Landau and Lifshitz 
(and appearing in one of their texts).  
This equation does admit neither preaccelerative nor postaccelerative
solutions, in common with many other replacements for the
Lorentz-Dirac equation proposed in the literature over the past sixty years.  
Also in common with most of these proposed replacements
is the fact that the Landau-Lifschitz proposal is  
an uncontrolled approximation to the Lorentz-Dirac equation.

A survey of such proposed equations (which unfortunately does not include
the proposal of Landau and Lifshitz) is given in \cite{parrottbook},
Section 5.7. 
Three of them are very similar in structure to the Landau-Lifshitz equation.
Their common features are that they are second order in
the particle's worldline (instead of third order as is the Lorentz-Dirac
equation) and their solutions reduce to inertial (i.e., zero acceleration)
motion when the external electromagnetic field vanishes. 
They are all obtained by approximating the acceleration in certain
terms of the Lorentz-Dirac equation by the Lorentz force.
{\em Any} equation with these features automatically eliminates
preacceleration, postacceleration and runaway solutions
for electromagnetic fields applied for only a finite time.  

These equations have been around for decades without achieving 
substantial acceptance.  Most been criticized in the literature
for predicting unphysical effects different from those noted above
for the Lorentz-Dirac equation.  

The Landau-Lifshitz 
equation has not escaped  criticism.  
For example, it predicts that a charged particle moving in a straight line 
in a uniform
electric field in the direction of that line
experiences {\em no} radiation reaction 
(and hence presumably doesn't radiate), a conclusion which some physicists
regard as contradicting physical observations of {\em bremsstralung}.
({\em Bremsstralung}, ``braking radiation'', is radiation observed
when energetic charged particles are slowed by collision with a target.) 

If Professor Rohrlich finds the Landau-Lifshitz equation such a compelling
resolution of the logical problems of classical electrodynamics, 
why didn't he mention this in his December, 2003 
report on my original paper \cite{parrottpaper}?
And why did he publish \cite{rohrlichpr} and  
\cite{rohrlichamj} advocating the DD equation?

Surely there must have been some aspect of the Landau-Lifschitz equation
which he found less than satisfactory.
Although Spohn's work was not published until 2000, after 
Rohrlich's DD equation proposals \cite{rohrlichpr} and \cite{rohrlichamj}, the Landau-Lifschitz equation
has been around for over forty years.  It is hard to imagine that 
Professor Rohrlich was unaware of it when he submitted 
his PRD paper \cite{rohrlichpr}, and he was certainly aware both of it
and of Spohn's work when he submitted his 2003 report on
the earlier version of my paper \cite{parrottpaper}. 
Indeed, Spohn's paper \cite{spohn}, published in 2000,
 specifically thanks him for calling the 
Landau-Lifschitz equation to his attention: 
\begin{quote}
``I am most grateful to Fritz Rohrlich for instructive discussions
and for the hint that the independently derived Eq.\ (8)
[the Landau-Lifschitz equation, which Spohn had obtained independently
as an uncontrolled approximation to the Lorentz-Dirac equation]
appeared in Landau and Lifschitz already a long time ago.''
\end{quote}
Why did Rohrlich's 2003 report on my paper \cite{parrottpaper} 
contain no mention of Spohn's 2000 ``solution'' (as claimed by Rohrlich,
not by Spohn) of the ``preacceleration problem''? 
Why did it give no  hint that he considered the DD equation
``no longer of physical interest''?

This question is not presented in a sarcastic or derisive way. 
Professor Rohrlich's latest report on my ``Comment'' paper  asks us to accept, 
{\em solely} on the strength
of his reputation, that Spohn's work solves 
the problems associated with the Lorentz-Dirac equation (a claim which
Spohn himself does not make), thereby making further research in this area
``no longer of physical interest''.
In assessing the validity of this claim, Rohrlich's
recent related statements, 
such as his 2003 report on my paper \cite{parrottpaper}, 
should be taken into consideration. 

Any advocacy of the Landau-Lifshitz equation ought to be based on
a detailed analysis of the predictions of that equation,
not on the fact that it is obtained as an uncontrolled approximation
to the Lorentz-Dirac equation.  Given the problems with the Lorentz-Dirac
equation noted above, the fact that some equation may approximate it
seems a questionable recommendation for that equation. 

Spohn's work neither solves any problem with the Lorentz-Dirac equation,
nor, in my view, provides convincing motivation for the
Landau-Lifschitz equation.  An appendix analyzes in some detail
what Spohn's mathematics actually establishes, and it it is quite
far from what Professor Rohrlich seems to believe.  
I don't want to become diverted here into an analysis of Spohn's work
because it is essentially irrelevant to whether a paper commenting
on Rohrlich's PRD \cite{rohrlichpr} proposal of the DD equation is justified.  

Rohrlich's PRD paper \cite{rohrlichpr}, makes  
claims for which there was never any real evidence. 
If these claims remain uncorrected,
many readers will uncritically accept them 
on the basis of his considerable reputation as an expert
on the foundations of classical electrodynamics, or on the basis
of a common perception that PRD is a reliable journal. 
Others with an interest in these problems may, like me,
invest months of hard mathematical work 
trying to resolve the validity
of these claims, work which is likely to be ultimately wasted.  
Scientific integrity demands that PRD open its pages to correction
of these misleading claims.  
\subsection{Appendix on the work of Spohn cited by Rohrlich} 

Professor Rohrlich's report on my ``Comment'' paper claims that
the paper is ``no longer of physical interest'' because
the problem which the DD equation was intended to solve has since
been solved by Spohn \cite{spohn}, and therefore renders 
the Comment paper (and, presumably, the DD equation)
``no longer of physical interest''. 

This appendix summarizes the part of  
the 2000 paper of Spohn \cite{spohn}
relevant to the claim that Spohn has somehow solved 
``the preacceleration problem''.
Rohrlich doesn't mention the numerous other problems with the 
Lorentz-Dirac equation, but his report gives the impression
that he thinks Spohn has solved those as well.

First of all, Spohn's paper \cite{spohn} 
does not claim to solve any ``preacceleration problem''.
Preacceleration is not even mentioned in this paper,
and nothing in the paper has anything to do with preacceleration.
Indeed, Spohn \cite{spohn} 
doesn't directly claim to solve {\em any} of the generally accepted
logical problems with the
Lorentz-Dirac equation.

However, it is not hard to imagine that a casual reader 
might get the incorrect impression
that he somehow disposes of the problem of runaway solutions
for the Lorentz-Dirac equation.
I shall now attempt to explain, without going into technical details,
what Spohn's mathematics actually does show.

The mathematics of his paper is based on 
1971 work of Fenichel on singular systems 
of ordinary differential equations \cite{fenichel} \cite{fenichel2}
as expounded in the monograph of C. Jones \cite{jones}.
Spohn's paper is sketchily  written,
and had I not been previously acquainted with Fenichel's nontrivial work,
Spohn's account would have been virtually incomprehensible to me. 
I would be surprised if many physicists are aquainted with Fenichel's highly
technical 
mathematics---Spohn's paper is the first time I have seen it mentioned
in the physics literature.  
I think that anyone {\em not} previously acquainted with Fenichel's work
would be unlikely to be able to read Spohn's paper with understanding. 
The issue here is whether Professor Rohrlich has the mathematical
background necessary to fully understand what Spohn has done,
and whether he has carefully checked Spohn's mathematics.  

I hope the independent referees will make an effort to determine this.  
If he continues to maintain that Spohn's work solves some fundamental
problem in electrodynamics, he should  be asked to respond to the
objections to follow.  I would be happly to write them out in greater
detail to facilitate such a response.

Now we begin the explanation of Spohn's application of Fenichel's
results to the Lorentz-Dirac equation. 
The Lorentz-Dirac equation for a particle of charge $q$ and mass $m$ 
can be written,
in the notation of my paper: 
$$
 \frac{du^i}{d\tau} = \frac{q}{m} {F^i}_\alpha u^\alpha + \epsilon 
\left[\frac{d^2 u^i}{d\tau^2} + \frac{du^\alpha}{d\tau}
\frac{du_\alpha}{d\tau} u^i
\right]
$$ 
where $\epsilon := 2 q^2/3m $ is a dimensionless parameter 
which is very small for real particles.
Note that for $\epsilon = 0$, the Lorentz-Dirac equation reduces
to the Lorentz equation, which describes charged particles 
neglecting effects of radiation.  
Note also that $\epsilon$ multiplies the term of highest differential
order, namely $d^2 u^i/d\tau^2$ of order 2, and that setting $\epsilon = 0$,
reduces the differential order of the equation from two to one. 
A differential equation containing a parameter $\epsilon$,
such that setting $\epsilon = 0$ changes the fundamental
nature of the equation (e.g., from order 2 to order 1),
is called a ``singular'' equation. 

For simplicity, we shall regard  
$F = F(\tau)$ as given, {\em a priori}, as a function of
proper time $\tau$.
For $\epsilon \neq 0$, the above form of the Lorentz-Dirac equation 
has a solution manifold 
of dimension eight, which is larger than the physically expected dimension
of four (which reduces to three after taking into account the 
fact that a four-velocity $u$ has to satisfy $u^\alpha u_\alpha = 1$).
Thus physically relevant 
solutions are expected to lie on an invariant submanifold 
of dimension three within the eight-dimensional manifold of all solutions;  
here ``invariant'' means invariant under the solution flow.

Fenichel established (under technical hypotheses)
the existence of an $\epsilon$-indexed 
family 
of invariant submanifolds $ M_\epsilon$, defined for sufficiently
small $\epsilon \geq 0$, which vary smoothly
with $\epsilon$.
It is trivial to obtain such a family for $\epsilon > 0$
(just choose a fixed initial acceleration $du/d\tau (0) = a_0$,
and parametrize the manifold by the initial velocities),
but if we insist on including $\epsilon = 0$, where the order of the 
equation abruptly changes, the problem becomes nontrivial.

However, the mathematically nontrivial problem of including 
$\epsilon = 0$ is of questionable relevance for physics.
In the real world, a particular particle of charge $q$ and mass $m$
is associated with a definite ``physical'' $\epsilon = 2q^2/3m$.
This physical $\epsilon$ can't be varied.  

The invariant submanifolds  $M_\epsilon$
which Fenichel obtained are called ``critical manifolds'' by Spohn.   
Spohn explains the terminology by a claim that all solutions off
the critical manifold must be runaway, 
so that all physically possible solutions must lie on the critical manifold. 
This claim is unproved in \cite{spohn} for the Lorentz-Dirac equation, 
though a ``toy'' model is presented to make it plausible by analogy.  

Next, Spohn presents an argument which claims to prove that
(under very special hypotheses given below, which Spohn does not state) 
solutions of the Lorentz-Dirac equation which {\em do}
lie on the critical submanifold are {\em never} runaway,
but instead have acceleration tending asymptotically to zero
in the infinite future (and past).
I believe this argument to be fundamentally incorrect.

For one thing, it contradicts Eliezer's Theorem stating that
in certain situations (which satisfy Spohn's hypotheses),
{\em all} solutions are runaway.  So, 
either Eliezer's theorem (whose proof has survived careful checks
over more than 60 years) must have an erroneous proof,
or Spohn's argument must be incorrect.

I am sure that it is Spohn who is incorrect because
\begin{enumerate}
\item
I can point out the precise places where his argument becomes invalid, and
\item
I have carefully checked the proof of Eliezer's Theorem. 
\end{enumerate}

Spohn's argument appears to be based on the {\em assumption},
that the particle's acceleration is bounded
on the critical manifold.    
If so, what his argument actually shows 
(under his other restrictive hypotheses)
is that {\em if} the particle's acceleration is bounded on the critical
manifold, then the acceleration tends asymptotically to zero. 
But this assumption rules out runaway solutions by hypothesis.

Another problem with the argument is that his ``energy balance''
equation (equation (6) of his paper), on which his argument is based,
 seems to require 
that the external field $F$ be of a very special form.
It seems to require the unstated assumption that there exists
a fixed Lorentz frame in which $F$ is derivable from a time-independent
scalar potential with zero vector potential.
This implies that $F$ is a static, pure elecric field in that frame.

For most electromagnetic fields 
(including all so-called ``radiation fields''),
it is impossible to choose a fixed Lorentz frame in which $F$ 
is a pure electric field.  A quick way to see this is to 
recall that the quantities 
\newcommand{\bE}{{\bf E}}
\newcommand{\bB}{{\bf B}}
$\bE^2 - \bB^2$ and $\bE\cdot \bB$, with $\bE$ the electric field 
and $\bB$ the magnetic field, are ``invariants'' in the sense that
they are independent of the Lorentz frame in which they are computed.
For ``radiation'' fields 
(examples of which are easy to write down explicitly), 
both of these quantities vanish identically,
so such a frame is impossible for nonzero radiation fields.  
Thus even if the other objections to  Spohn's argument could be overcome,
the resulting conclusion (that solutions on the critical manifold
vanish asymptotically) would be too special to be considered a solution
of any fundamental problem with the logical foundations of electrodynamics. 

The only other result in Spohn \cite{spohn} of interest here is that he 
calculates an approximate equation for the critical manifolds,
valid to first order in $\epsilon$.  This is then used to obtain
the Landau-Lifschitz equation as an uncontrolled approximation
to the Lorentz-Dirac equation.

I don't see why obtaining Landau-Lifshitz as an uncontrolled approximation
to Lorentz-Dirac using all this fancy mathematics is any better
than obtaining it as an uncontrolled approximation in simpler ways.
Moreover, since nobody believes the consequences of the Lorentz-Dirac
equation listed in the Introduction (e.g., that in some circumstances,
unlike charges repel instead of attract), why should one have any
confidence in an equation obtained as an approximation to the 
Lorentz-Dirac equation?  And why should it be considered a definitive
solution to any of the logical problems of classical electrodynamics?
\section{Appendix 3:  
Remarks and corrections concerning the reply to Rohrlich's report}
For historical accuracy, I included the reply to Professor Rohrlich's
report verbatim as it was submitted to {\em Phys. Rev. D} (PRD),
including some insignificant typos (e.g., ``happly'' for ``happy").  
This section corrects some 
inadvertent misstatements (luckily minor) 
in that reply and adds some comments.

After PRD refused to publish the ``Comment'' paper, 
I wrote Dr.~Spohn
to let him know that I intended to include the reply as an appendix
to the copy in the ArXiv.
The message offered to correct any inaccuracies and to include a reply
from him should he care to furnish one. 
I thank him for subsequent correspondence 
which deepened my understanding
of the intent, assumptions, and conclusions of his approach to
the Lorentz-Dirac (LD) equation.
I also thank him for 
sending me a reply in LaTeX format, which comprises
Appendix~4. 
\begin{enumerate}
\item
The constant $\epsilon := 2q^2/3m$ in the Lorentz-Dirac equation
on page 7 is not dimensionless as stated---it has the dimensions of time. 
This does not affect any of the conclusions.
\item
I objected to Rohrlich's statement that Spohn's paper \cite{spohn}
``showed'' (i.e., proved) that ``the Lorentz-Dirac equation
has in its solution space a ``critical manifold'' to which
all physical solutions must be restricted''.
Although \cite{spohn} does give the impression of claiming this
(it certainly isn't proved there),
Dr.\ Spohn informs me that this is a conjecture rather than a theorem. 
\item
Dr.\ Spohn confirms that his ``energy balance'' equation 
(which is the basis of his argument) requires the assumption
(unstated in \cite{spohn})
that there exists a Lorentz frame in which the potentials are static
(i.e., time -independent), and hence the field $F$ is static.  
\s
In guessing what additional conditions might be assumed by \cite{spohn}
to justify his energy balance equation, I guessed a condition which 
was too strong---I guessed that he was assuming that $F$ was 
static with vanishing magnetic field. 
\newcommand{\brr}{{\bf r}}
\newcommand{\bA}{{\bf A}}
\newcommand{\bbeta}{{\bf \beta}}
\newcommand{\bx}{{\bf x}}
\newcommand{\bnabla}{{\bf \nabla}}
\newcommand{\bP}{{\bf P}}
\newcommand{\bQ}{{\bf Q}}
\newcommand{\bk}{{\bf k}}
\s
Assuming only his weaker condition (static $F$), my previous argument that  
it is not generally possible to find such a Lorentz frame
no longer suffices.  However, it is still true that such a Lorentz
frame need not exist.  One class of fields for which such a Lorentz
frame cannot exist consists of fields which in three-dimensional notation
(relative to some given Lorentz frame, with three-vectors in boldface) 
have electric and magnetic fields $\bE$ and $\bB$
of the form  
$$
\bE (x) =  \bP e^{ikx}, \q \bB (x) = \bQ e^{ikx}
\q.
$$
Here $\bP$ and $\bQ$ are constant three-vectors 
which are orthogonal
to each other and to the space part $\bk$ of the four-vector
$k = (k^0, \bk) $, and  for $x = (x^0, \bx)$, 
$kx$ denotes Lorentz inner product $kx := x^0k^0 - \bx \cdot \bx$.
These will satisfy the free-space Maxwell equations if $k^0 = |\bk|$
and $\bk \times \bP = - k^0 \bQ$.
Although these are written as complex fields for algebraic simplicity,
similar real fields for which the argument to follow is valid can be obtained
by taking real or imaginary parts.
\s
For $k \neq 0$, these fields are obviously not static (relative to
the given Lorentz frame).  Moreover, the standard formulas for transforming
them to a new Lorentz frame yield fields of the same form
(with generally different $\bP$ and $\bQ$), so they cannot  
be static in the new frame, either.
\item
Dr.\ Spohn has confirmed that, as I had guessed, 
the argument of \cite{spohn} does assume that the acceleration
is bounded on the critical manifold.
He says that this follows from something in Sakamoto's paper
\cite{sakamoto}. 
(Neither the assumption nor its justification by Sakamoto
are mentioned in \cite{spohn}.)
Since the nearest copy of that paper is probably 200 miles away,
I am not in a position to confirm this.
\s
This argument also requires the unstated assumption that 
the velocity on the critical manifold is bounded away from the speed 
of light---I do not know if this also derives from Sakamoto. 
\item
The argument of \cite{spohn} requires that the  
potentials be bounded on the particle's worldline,
and Eliezer's theorem does guarantee this for a Coulomb field. 
This is why I thought that Spohn's results must conflict
with Eliezer's theorem for a Coulomb field.
However, Dr.\ Spohn informs me that another unstated assumption
of \cite{spohn} is that the potentials be globally bounded 
(not just bounded on the worldline).
This added assumption removes the formal contradiction
between Eliezer's theorem and Spohn's work. 
\s
After learning this, I was still skeptical because there
are forms of Eliezer's theorem which apply to 
the potential
of a Coulomb field which is cut off at both large  
and small distances from its source
(e.g., the scalar electric field $E = 1/r^2$ for $r_1 \leq r \leq r_0$
and vanishing elsewhere). 
(The cutoff can be ``smoothed'' near $r = r_1$ and $r = r_0$, if desired,
to obtain a $C^\infty$ field.)  
It seemed to me that the conclusion of \cite{spohn}
that solutions on the critical manifold tended asymptotically
to zero might be in conflict with such variants of Eliezer's Theorem. 
\s 
One of  these
variants concludes that for a charged particle
which enters such a cutoff Coulomb field
with sufficiently small velocity, 
all solutions of the Lorentz-Dirac   
equation are runaway.
However, it requires the additional hypothesis (not required by
Eliezer's original theorems) that the initial acceleration 
(the acceleration when the particle enters the field) must vanish. 
This hypothesis is equivalent to assuming that there 
is no preacceleration.  
\s
Under the hypothesis that preacceleration is impossible,
the results of \cite{spohn} 
seem in direct logical contradiction
to this variant form of Eliezer's theorem. 
Put differently, 
{\em all} solutions of the Lorentz-Dirac equation (for a cutoff
Coulomb field and sufficiently small initial velocity) 
are either runaway or preaccelerative. 
In particular, assuming Spohn's conclusion \cite{spohn}
that solutions on his critical manifold are asymptotic to zero
in the future (hence not runaway), these solutions must be preaccelerative. 
\s
Thus it is hard to see what could justify Professor Rohrlich's claim
that \cite{spohn} solves ``the preacceleration problem''.  
(It should be emphasized that Dr.\ Spohn has never made any such claim,
to my knowledge, either in print or in correspondence with me.)
\s
The precise statements and proofs of these variant forms
of Eliezer's theorem can be found in www.arxiv.org/math-ph/0505042. 
\end{enumerate} 
\section{Appendix 4:  Dr. Spohn's reply}
\newcommand{\epsi}{\varepsilon} 

\noindent Comment by\bigskip\\ 
Herbert Spohn\\
Physik Department and Zentrum Mathematik, TU M\"{u}nchen\bigskip

Dr. Parrott kindly asked me to add a comment to the exchange of
opinions. My full view on radiative friction is explained in my
recent book {\it Dynamics of Charged Particles and Their Radiation
Field, Cambridge University Press, 2004}. Since the exchange of
opinions is
focused  on my 2000 Europhysics Letter, I briefly explain
what I did and did not achieve in this contribution.\smallskip\\
(i) I point out that the Lorentz-Dirac equation is a singularly
perturbed differential equation, a topic on which considerable
mathematical knowledge has accumulated. The same observation was
stressed by Galgani and coworkers already in 1995. His and more recent studies 
from this point of view are cited in my book.\smallskip\\
(ii) Locally  in space-time a singularly perturbed differential
equation has a center (or critical) manifold of ``slow'' motion. Globally this
manifold may have a complicated structure. I argue that the physical
solutions of the Lorentz-Dirac equation must be on its center 
manifold, thereby improving on Dirac's asymptotic
condition.\smallskip\\
(iii) If one expands the center manifold in the small
parameter, then to first order the motion on the center manifold is
governed  by the Landau-Lifshitz equation for
radiative friction, which provides a more systematic understanding 
for a otherwise seemingly {\it ad hoc} procedure.\smallskip\\
(iv) One point in the discussion with Dr. Parrott, also in  relation to Eliezer's
theorem, is the structure of the center manifold in concrete cases
like the motion in an attractive Coulomb potential. This issue leads to the
question under what conditions the Landau-Lifshitz equation
 is a good approximation to the true center manifold motion. 
In my opinion such points can be clarified only
through more detailed investigations.
Considered as a low dimensional dynamical system the
Lorentz-Dirac equation has many still unexplored features. 
\bigskip\\
Herbert Spohn, May 20, 2005
\section{Final  thoughts}
The mathematical analysis of this paper and its precursor 
\cite{parrottpaper} took a few months.
The time from submission to final rejection was about 
three years.  In the process, several hundred pages of correspondence
were generated, probably requiring more time than did the initial
mathematical analysis.
It has been a frustrating and exhausting process, 
and I am relieved that it is finally over. 

Before putting this work to final rest in the ArXiv,
I have been asking myself what can be learned from this experience.
The subsections below summarize my conclusions and raise some
questions which I hope others in this field will consider. 
\subsection{Editorial standards of {\em Phys. Rev. D} (PRD)}
This ``Comment'' paper was apparently rejected solely because
an anonymous referee deems its subject 
(proposed equations of motion for classical charged particles)
as ``not of physical interest''. 
Since its subject 
is identical to the subject of Rohrlich's recent PRD paper \cite{rohrlichpr}
on which it comments, this seems to put PRD in the strange 
position of affirming that Rohrlich's paper \cite{rohrlichpr},
which they recently published, 
was so uninteresting that comments on it are superfluous. 
I wouldn't have believed that the editors of PRD could maintain this 
with a straight face.

The anonymous referee's characterization may be a defensible personal
opinion, but  
it is certainly not  universally accepted.  
After  PRD published
Rohrlich's \cite{rohrlichpr} in 1999,
it has continued publishing papers on proposed equations
of motion for classical charged particles, about one paper a year
by my count.

Less than a year ago (Fall, 2004), 
Cambridge University Press published a book
by H. Spohn \cite{spohnbook} 
which prominently emphasizes this subject. 
It presents his method of obtaining the Landau-Lifshitz  equation
from the Lorentz-Dirac equation, and it applies the Landau-Lifshitz 
equation to draw conclusions about the behavior of an electron
in a Penning trap. 

While it is surely true that not all physicists are interested 
in this subject, it is also unquestionable that does exist 
a substantial
community which is interested.  
Some in this community hope that
progress in constructing a consistent classical electrodynamics
might lead to a consistent quantum electrodynamics.  Others
hope that new technologies such as Penning traps might make it possible
to extract useful predictions from classical equations of motion.  

It would be a legitimate editorial decision to refuse to accept 
all papers dealing with equations of motion for classical charged particles. 
However, that decision should be made
on the editorial level, 
not arbitrarily by referees. 
And, if such a policy is adopted, it should be made public and uniformly 
enforced.

Before I prepared the ``Comment'' paper, I wrote the Editor of PRD,
\newline
D.~Nordstrom, asking
specifically if PRD would consider a ``Comment'' paper based
on the rejected \cite{parrottpaper}.  He did not reply.  
If the subject of the paper were, {\em a priori}, unacceptable to PRD, 
the editors should have 
informed me of that when I asked. 
\subsection{Unreliability of the physics literature}
The process of researching my book \cite{parrottbook} 
made me painfully aware of the general unreliability of the 
literature on classical relativistic dynamics.
My experience has been that
something like a third to a half of all published papers in this field
contain errors sufficiently serious to potentially invalidate
some of their main conclusions. 

The likely explanation is that few papers are carefully read by
their referees.  The reasons are easy to guess---refereeing
is tedious volunteer work which confers inadequate professional 
rewards.  Nevertheless, this is not a necessary state of affairs.
For example, the mathematical literature is much more reliable
than the physics literature despite a similar refereeing
system.

The unreliability of the physics literature 
forces serious readers to check all calculations
in extraordinary detail. 
Since many of the calculations in this field are very tedious,
this creates an enormous waste of time.  Every serious reader
has to duplicate this work,
to be reasonably sure of the correctness
of a paper.    

To make matters worse, discovered errors are seldom corrected.
The experience of this ``Comment'' paper clearly shows how 
difficult it can be to get correction of errors into the literature.

For example, the elementary mathematical errors of Moniz and Sharp's
PRD analysis \cite{ms} of the nonrelativistic DD equation have gone
uncorrected for almost 30 years.
%
Rohrlich's recent PRD article \cite{rohrlichpr} 
relied on Moniz and  Sharp's incorrect analysis to justify, 
by analogy, 
its questionable claims about the relativistic DD equation. 
Given that PRD is evidently unwilling to correct
such misleading claims and outright errors, 
it is easy to imagine that 
other work based on these errors may continue to appear. 

The standards of PRD are, unfortunately, not atypical among physics 
journals.  Given this, little can be done 
about the unreliability of the literature within the present system.  

However, the ArXiv could be a powerful 
tool for reform because articles posted there are permanently accesible
and subject to peer review.  Though the peer review is informal,
it could easily become more reliable 
than the current refereeing system for journals.  
I would place more weight on a citation of an ArXiv paper
by someone whom I knew to be competent than on acceptance
in a journal like PRD which publishes much erroneous work.
\subsection{The role of the ArXiv}
Given this experience,
I will never submit another paper to PRD.
Indeed, I doubt that I shall submit any of my future work for publication
to physics journals.  As a retired person, publication
gives no professional benefit, and 
it is highly unpleasant to have to deal with incompetent referees 
and journals which lack scientific integrity.  
Instead, I plan to post my future work in the ArXiv, and this raises
another issue.

I am uneasy about recent changes in ArXiv policies which make it 
more difficult to place papers there.  Some people are required
to find an ``endorser'' for any paper which they wish to post.
I am not so required because I was grandfathered
in.  But if I did need to find an  endorser, 
it might not be easy, and I might not want to try.  
For example, I am not sure that I could find an endorser
for this submission, and I probably wouldn't try.

The endorsement system is apparently meant to eliminate
``crackpot'' papers.  Unfortunately, 
a likely side effect will be
to effectively bar useful work from the ArXiv.  

It is not clear to me that significant harm is done
by letting a few misguided individuals post papers which
some might regard as ``crackpot''. 
Also, the line between ``crackpot''
and ``brilliant but inarticulate'' is not always clear. 
Cases of obvious abuse (e.g., if someone attempts to
``publish'' a  novel in gr-qc) could be handled by 
other means.%
\footnote{
For example, users of the ArXiv could be asked to inform its
administration of any obviously inappropriate submissions,
which could be removed by hand.  
Repeat abusers could be barred from posting anything.
There will always be a few problems which cannot be mechanically
weeded out, but a few simple rules such as those suggested 
should cover the vast majority of abuses.  
And, one would not expect a great number of abuses anyway,
given the nature of the ArXiv.
} 
It does seem clear that if policies result in 
potentially useful work not being posted, 
the entire community loses. 

Scientific history is full of instances of work 
which could not find proper publication in its time 
but which later became generally recognized as valuable.  
So long as the ArXiv is partially supported by public funds,
the public should require that its policies enhance,
rather than limit, free scientific expression. 

\end{document}